# A conditionally exactly solvable generalization of the inverse square root potential

A.M. Ishkhanyan[1,2,3]
[1]Institute for Physical Research, NAS of Armenia, Ashtarak 0203, Armenia
[2]Armenian State Pedagogical University, Yerevan 0010, Armenia
[3]Institute of Physics and Technology, National Research Tomsk Polytechnic University, Tomsk 634050, Russia

We present a conditionally exactly solvable singular potential for the one-dimensional Schrödinger equation which involves the exactly solvable inverse square root potential. Each of the two fundamental solutions that compose the general solution of the problem is given by a linear combination with non-constant coefficients of two confluent hypergeometric functions. Discussing the bound-state wave functions vanishing both at infinity and in the origin, we derive the exact equation for the energy spectrum which is written using two Hermite functions of non-integer order. In specific auxiliary variables this equation becomes a mathematical equation that does not refer to a specific physical context discussed. In the two-dimensional space of these auxiliary variables the roots of this equation draw a countable infinite set of open curves with hyperbolic asymptotes. We present an analytic description of these curves by a transcendental algebraic equation for the involved variables. The intersections of the curves thus constructed with a certain cubic curve provide a highly accurate description of the energy spectrum.

## 1. Introduction

In this brief communication we introduce a conditionally exactly solvable potential for the one-dimensional Schrödinger equation which is a generalization of the recently presented exactly solvable inverse square root potential [1]. The term "conditionally exactly solvable" is conventionally used as referring to the potentials for which a parameter is fixed to a certain value [2-7]. Well known examples of this kind of potentials are the two Stillinger potentials [2] that have been discussed by many authors (see, e.g., [3]). Many other potentials can be found in [4-7] and references therein.

The potential that we introduce, however, does not directly meet the given definition of conditionally exactly solvable potentials because none of the parameters of this potential is fixed. Rather, two of them depend on each other. To be specific, here is the potential:

$$V = V_0 + \frac{V_1}{\sqrt{x}} + \frac{8mV_2^2/\hbar^2}{x} + \frac{V_2}{x^{3/2}}, \qquad (1)$$

$x \in (0, +\infty)$. This potential is straightforwardly obtained by inspecting the equations (A.7) and (A.8) of the Appendix of [1] for the case when the parameters involved in the trial potential (A.8) are allowed to depend on each other. The general solution of the Schrödinger equation for the potential is then directly written following the lines of that Appendix.

As it is seen, the strengths of the $1/x$ and $1/x^{3/2}$ terms of the potential are not fixed. However, we see that these parameters cannot be varied independently. This is why the potential is not, in the conventional sense of the term, a strictly "conditionally exactly solvable" one. On the other hand, it is also intuitively understood that the potential cannot be referred to as "exactly solvable". To our opinion, the situation is resolved if we accept the convention to apply the term "exactly solvable" exclusively to the potentials that are proportional to a parameter, say $\mu$, and have a shape $S(x)$ that is independent of that parameter, that is, to the potentials that can be written (in a coordinate system) as $V = \mu \cdot S(x)$ with $S \neq S(\mu)$. Additionally, both the parameter $\mu$ and the potential shape $S(x)$ are assumed energy-independent. This definition puts the potential (1) out of the family of the "exactly solvable" ones. Then, what term better fits this potential? Our suggestion is to apply the term "conditionally exactly solvable" thus implying that the word "conditional" is used in the direct sense of the word as a restriction imposed on the parameters of the potential. Broadly speaking, any restriction, not only that a parameter is fixed to a certain value. The latter potentials can then be referred to as "fixed-parameter" potentials. Perhaps, this makes the term "conditionally exactly solvable" more self-consistent.

Thus, we refer to the potential (1) as conditionally solvable. We present the general solution of the one-dimensional stationary Schrödinger equation:

$$\frac{d^2\psi}{dx^2} + \frac{2m}{\hbar^2}\left(E - V(x)\right)\psi = 0, \qquad (1)$$

for this potential and explore the properties of the solution which is written in terms of the confluent hypergeometric functions. A peculiarity of the solution that is worth to be mentioned is that each of the two fundamental solutions that compose the general solution of the problem is written as a linear combination with non-constant coefficients of two confluent hypergeometric functions. It is convenient to write one of the functions involved in the fundamental solutions as a Hermite function (generally, of non-integer order).

We derive the exact equation for the energy spectrum for the bound states and explore the properties of the spectrum. The spectrum equation which involves two Hermite functions is further rewritten as a mathematical equation that does not refer to a specific physical context we discuss. In the two-dimensional space of the introduced auxiliary variables the roots of this equation draw a countable infinite set of slightly asymmetric open curves with hyperbolic asymptotes. We derive a highly accurate description of the curves trough a transcendental algebraic equation for the involved variables. The intersections of these curves with a certain cubic curve present a highly accurate description of the energy spectrum.

## 2. General solution and energy spectrum

The straightforward result immediately achieved by following the lines of the Appendix of [1] if the equation (A.7) is inspected for the case of dependent parameters involved in the trial potential (A.8) is that the general solution of the Schrödinger equation for arbitrary (real or complex) parameters $V_{0,1,2}$ involved in potential (1) is written as

$$\psi(x) = e^{\frac{\delta z^2}{4} + \frac{\gamma z}{2}} \frac{du}{dz}, \qquad (3)$$

with $\quad u = e^{-\frac{\delta z^2}{2} + \frac{\alpha z}{\delta} - \gamma z} \left( c_1 H_a \left( \frac{\delta(\gamma + \delta z) - 2\alpha}{\sqrt{2}\delta^{3/2}} \right) + c_2 \cdot {}_1F_1 \left( -\frac{a}{2}; \frac{1}{2}; \frac{(\delta(\gamma + \delta z) - 2\alpha)^2}{2\delta^3} \right) \right), \qquad (4)$

where $z = \sqrt{2x}$, $c_{1,2}$ are arbitrary constants, $H_a$ is the Hermite function, ${}_1F_1$ is the Kummer confluent hypergeometric function, and the involved parameters are given as

$$a = \frac{\alpha(\alpha - \gamma\delta)}{\delta^3} - 1, \quad \alpha = \frac{\gamma\delta}{2} - \frac{2\sqrt{2}mV_1}{\hbar^2}, \qquad (5)$$

$$\gamma = \frac{8\sqrt{2}mV_2}{\hbar^2}, \quad \delta = \sqrt{\frac{8m(-E + V_0)}{\hbar^2}}. \qquad (6)$$

Potential (1) is a member of the first family of the bi-confluent Heun potentials [8-10] which include many important potentials intensively applied in the past. For $V_2 = 0$ the potential reduces to an exactly solvable potential, the inverse square root potential $V_1/\sqrt{x}$, the solution for which has been recently presented in [1]. For $V_1 = 0$ the potential reproduces the recent conditionally exactly solvable result by López-Ortega [11]. We note that potential was first introduced by Exton in [12].

The inspection of the potential shows that for a certain variation region of the involved parameters it supports bound states. It is understood that the bound states are possible if $V_1 < 0$. The behavior of the potential for $V_1 = -10$ and different positive and negative values of the parameter $V_2$ is shown in figure 1. The attractive inverse square root potential for which $V_2 = 0$ and $V_1 < 0$ does support bound states. These states have been discussed in [1]. The bound states for the potential (1) for the general case $V_2 \neq 0$ are derived essentially by the same approach as the one applied in [1]. Indeed, by demanding the wave function to vanish in the origin and at infinity [3], one readily reveals that $c_2 = 0$ and the *exact* energy spectrum equation is written as

$$H_{a-1}\left( \frac{\gamma\delta - 2\alpha}{\sqrt{2}\delta^{3/2}} \right) + \frac{\alpha - \gamma\delta}{\sqrt{2}a\delta^{3/2}} H_a\left( \frac{\gamma\delta - 2\alpha}{\sqrt{2}\delta^{3/2}} \right) = 0. \qquad (7)$$

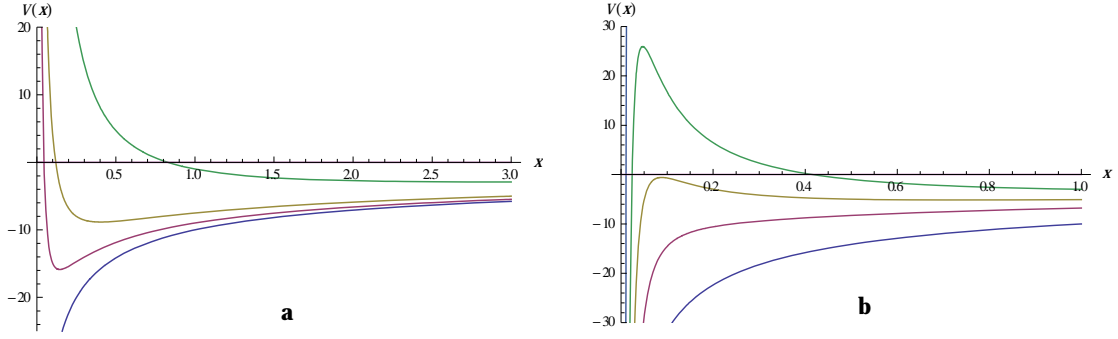

Fig.1 Potential (1) for (**a**) $V_1 = -10$ and positive $V_2 = 0, 0.3, 0.5, 1$ (from down to top) and (**b**) $V_1 = -10$ and negative $V_2 = 0, -0.7, -0.85, -1$ (from down to top) in units $m = \hbar = 1$.

In terms of the auxiliary variables

$$v = \frac{2\sqrt{2}mV_1}{\delta^{3/2}\hbar^2}, \quad w = \frac{4\sqrt{2}mV_2}{\delta^{1/2}\hbar^2} \tag{8}$$

this equation becomes a mathematical object that does not refer to a specific physical context:

$$H_{v^2-w^2}\left(\sqrt{2}v\right) - \sqrt{2}(v-w)H_{v^2-w^2-1}\left(\sqrt{2}v\right) = 0. \tag{9}$$

In the two-dimensional space of the involved variables $(w,u)$ the roots of this equation draw a countable infinite set of curves shown in Fig. 2. At $v = w$ the left-hand side of the equation is simplified to $H_0(\sqrt{2}v) = 1 > 0$, and for $v = -w$ it equals $1 + \sqrt{2\pi} we^{2w^2}(\text{erf}(\sqrt{2}w)+1) > 0$. Hence, the curves presenting the roots of equation (9) are bounded by the lines $v = \pm w$ shown in the figure. As the curve number $n$ increases the curves more and more resemble an origin-centered south-opening hyperbola given by the equation $v^2 = w^2 + n + c_0$, $c_0 < 0$.

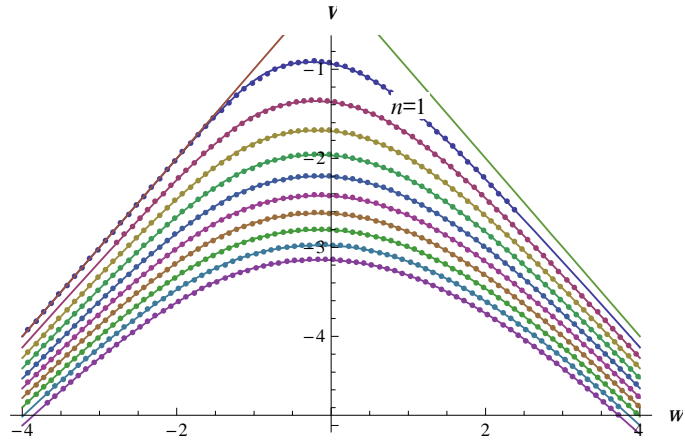

Fig.2 The first eight curves $v = v(w)$ presenting the roots of equation (9) (from top to down $n = 1, 2, ..., 8$). Filled circles - exact numerical result, solid curves - approximation (10).

Exploring equation (9) in detail, we divide it by $H_{v^2-w^2}(\sqrt{2}v)$ and use the known recurrence relations and series expansions for the involved Hermite functions [13] to get the following key result for the roots of the equation:

$$v^2 \approx w^2 + n - \frac{1}{2} + \frac{1}{2}\tanh\left(1 + \frac{n+2}{n+1}w\right), \quad n = 1, 2, 3, \ldots. \quad (10)$$

The accuracy of this approximation is demonstrated in figure 2, where the filled circles indicate the exact numerical values and the solid curves present the derived approximation. It is seen that this is a highly accurate approximation. For all curves it everywhere provides the values of the roots with relative error of the order or less than $10^{-4}$.

Returning back to the physical problem at hand, we note that it follows from the definitions (8) that the introduced quantities $v, w$ obey the relation

$$v = \frac{\hbar^4 V_1}{64 m^2 V_2^3} w^3. \quad (11)$$

In the $(w, v)$ parameter plane this equation draws a cubic curve the intersections of which with the curves presenting the roots of equation (10) determine the energy eigenvalues for the bound states for the potential (1). A peculiarity that is worth to be mentioned here is that the resultant bound state wave functions are not quasi polynomials because the orders of the involved Hermite functions are not integers.

Thus, with the definitions (8) equation (10) presents the equation for the spectrum (the parameter that introduces the energy into the equation is $\delta = \sqrt{8m(-E+V_0)}/\hbar$). This is, however, a transcendental equation the solution of which for the general case of non-zero $V_2$ is not known. To treat the equation, we first note that the hyperbolic-tangent term is small. Then, one could try to construct a reasonable first approximation valid for at least small $V_2$ by neglecting the $V_2$-dependent $w$-term in the argument of the hyperbolic-tangent term. However, in doing this we get a cubic equation for $\delta$ the treatment of which is still complicated because Cardano's cubic formula is cumbersome. One may use the trigonometric or hyperbolic methods to determine the real root of the resultant cubic equation; however, we then will still remain in the limits of the first approximation. A simpler and more advanced approach is to apply the successive iterations for the quantity $n_{eff} \equiv v^2$ in terms of which the energy spectrum for all orders according to the first equation (8) is written as

$$E_n = V_0 - \frac{1}{2}\left(\frac{mV_1^4}{\hbar^2}\right)^{1/3} \frac{1}{n_{eff}^{2/3}}. \quad (12)$$

Indeed, by taking $w^{(0)} = 0$ (that is $V_2 = 0$) as the starting iteration, for the zero-order approximation we immediately get

$$n_{eff}^{(0)} = n - \frac{1}{2} + \frac{\tanh(1)}{2} \tag{13}$$

and for the first approximation the result reads

$$n_{eff}^{(1)} = n - \frac{1}{2} + 16 V_2^2 \left( \frac{m^4 n}{\hbar^8 V_1^2} \right)^{1/3} + \frac{1}{2} \tanh\left( 1 + \frac{2+n}{1+n} 4 V_2 \left( \frac{m^4 n}{\hbar^8 V_1^2} \right)^{1/6} \right), \tag{14}$$

where in the $V_2$-proportional terms we have neglected, compared with $n$, the non-essential constant $c_0 = (-1 + \tanh(1))/2 \approx -0.119$.

As seen, the zero-order approximation does not depend on $V_2$. Hence, it corresponds to the pure inverse square root potential and thus may be applicable only in the very close vicinity of the point $V_2 = 0$. Even there, with the effective $n_{eff}^{(0)}$ given by equation (13), this result is less accurate than the one with $c_0 = -1/(2\pi) \approx -0.159$ derived in [1]. However, by inspecting equation (14), it is understood that in the first approximation this constant does not cause any essential variation so that for all orders $n \geq 1$ it can be neglected in the successive approximations without a noticeable loss in the accuracy.

As regards the result (14) of the first iteration, this is a reasonable approximation which is rather accurate for small $V_2$ and it is qualitatively correct everywhere. Hence, this is an approximation uniformly applicable for the whole variation range of $V_2$. Indeed, though inaccurate for large $V_2$, it provides quickly converging next approximations that are very accurate everywhere. Already the third iteration produces a result that is practically indistinguishable from the exact result in the whole variation range of $V_2$ (figure 3).

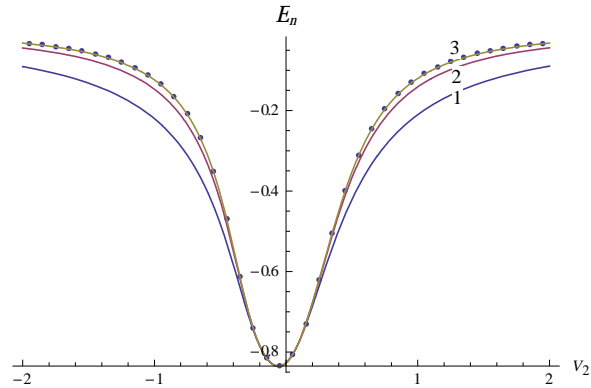

Fig.3 The first, second and third successive approximations for the second energy level $n = 2$ versus $V_2$ for $V_1 = -2$ (in units $m = \hbar = 1$). Filled circles indicate the exact numerical result and the solid curves present the successive approximations.

Then, by inspecting the result of the first three successive iterations, we conclude that the limiting result of the whole iteration is well described by an expansion of the form

$$n_{eff} = n + a_0 + a_1 n^{1/3} + a_2 n^{-1/3} + a_3 n^{-2/3} + \ldots. \tag{15}$$

This can also be shown analytically by applying the Banach fixed point theorem [14]. The rigorous numerical testing reveals that such an expansion provides a highly accurate result. Everywhere in the variation region of the parameters $V_{1,2}$ the first few terms provide the spectrum with the relative error of the order of or less than a few fractions of a percent.

This conclusion states that the energy spectrum approximately obeys the functional dependence $E_n \sim (n + a_0 + a_1 n^{1/3})^{-2/3}$ which much resembles the structure of the energy spectrum of the inverse square root potential [1]. The three-dimensional surface plot of the first two energy levels versus $V_{1,2}$ is shown in fig. 4. As seen, the effect of the conditionally exactly solvable component involving the $1/x$ and $1/x^{3/2}$ terms on the pure inverse square root potential's spectrum $E_n = E_1(n - 1/(2\pi))^{-2/3}$ consists in raising the energy levels (up to the limiting level $E = 0$ observed at large $V_2$) in both negative and positive $V_2$ cases. For large $V_1$, however, the effect is negligible. We thus conclude that the bound states are mostly conditioned, as expected, by the long-range component of the potential. This is well seen from equation (12) which resembles the structure of the spectrum of the inverse square root potential [1]. The effect of the short-range terms proportional to $1/x$ and $1/x^{3/2}$ is mainly in the modification of the effective level number $n_{eff} = n_{eff}(n)$ involved in equation (12).

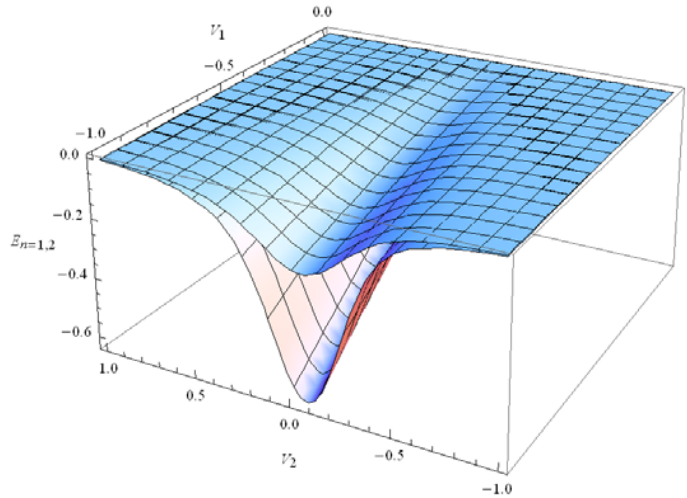

Fig.4 The first two energy levels versus $V_{1,2}$ ($m = \hbar = 1$).

## 3. Discussion

Thus, we have presented a conditionally exactly solvable potential that is the sum of the exactly solvable inverse square root potential $V_0/x^{1/2}$ [1] and the conditionally exactly solvable potential involving inverse power terms $1/x$ and $1/x^{3/2}$ [12]. This potential may model different physical situations. For instance, the $1/x^{1/2}$ singularity arises from a gate voltage at the edge of a graphene nanoribbon [15]. Since the quantum dynamics of electrons in graphene is described by a 1D Dirac equation, this is equivalent to solving the one-dimensional Schrodinger equation with a potential involving $1/x^{1/2}$, $1/x$ and $1/x^{3/2}$ terms.

We have presented the general solution of the one-dimensional Schrödinger equation and have discussed the bound states vanishing both at infinity and in the origin. We have derived the exact equation for the energy spectrum which involves two Hermite functions of non integer order. We have shown that the roots of this equation are rigorously described by a transcendental algebraic equation which provides a highly accurate description of the energy spectrum. Inspecting the result, we see that the spectrum for the particular potential much resembles that for the inverse square root potential. Besides, as in the case of the inverse square root potential, the bound state wave functions are not quasi polynomials.

We recall that the potential we have presented was first introduced by Exton, who constructed the solution of the Schrödinger equation for this potential as power series (note that the printed solution contains some misprints) [12]. Furthermore, Exton discussed the eigenvalue problem applying the polynomial reductions of his series. By a straightforward inspection with appropriate change of notations, one reveals that Exton's result is reproduced by the above eigenvalue approximation (10) if the hyperbolic tangent term there is replaced by its asymptotes at large $w$, that is by $\pm 1$. However, the polynomial solutions produce wave-functions that do not vanish in the origin. It is known that this is not acceptable physically. For a detailed discussion of this point, initiated by a paper by Souza Dutra [16], as well as for the correct boundary conditions that should be applied, see, for instance, [17-18].

We would like to conclude by recalling that the presented potential belongs to the class of the bi-confluent Heun potentials [8-10]. It is interesting that the potential is derived starting from an equation obeyed by a function involving the first derivative of the tri-confluent Heun function [19,20]. The application of the equations obeyed by the derivatives of the five Heun equations to the Schrödinger equation as well as to other analogous quantum or classical equations is rather productive. Supporting this observation is the derivation of several new exactly or conditionally exactly solvable potentials recently presented by several authors for different equations including the relativistic Dirac and Klein-Gordon equations [21-24]. Based upon the recent experience in treating the quantum two-state problem by this approach [25-27] we may expect more such results if a systematic study is undertaken. A last remark is that one may go even further by considering equations more general then the Heun equations. Examples of such equations [28] and their applications for solving the Schrödinger equation are known already from the early days of quantum mechanics [29] and have been recently discussed by several authors (see, e.g., [30]).

## Acknowledgments

I am very thankful to the referee who brought to my attention the important contribution by H. Exton, Ref. [12]. This research has been conducted within the scope of the International Associated Laboratory IRMAS (CNRS-France & SCS-Armenia). The work has been supported by the Armenian State Committee of Science (SCS Grant No. 15T-1C323) and the project "Leading Russian Research Universities" (Grant No. FTI_24_2016 of the Tomsk Polytechnic University).